\documentclass[prl,longbibliography,twocolumn,superscriptaddress,amsmath,amssymb,verbatim]{revtex4-1}

\usepackage{graphicx}
\usepackage{lmodern}
\usepackage{epstopdf}
\usepackage[colorlinks=true,citecolor=blue,linkcolor=blue,urlcolor=blue,bookmarks=true]{hyperref}
\usepackage{xcolor}
\definecolor{green}{rgb}{0.1, 0.6, 0.3} 
\newcommand{\red}{\color{black}}

\newcommand{\YCu}{YCu$_3$(OH)$_6$Cl$_3$}

\begin{document}

\preprint{APS/123-QED}

\title{Coexistence of magnetic order and persistent spin dynamics in a quantum kagome antiferromagnet with no intersite mixing}

\author{A.\,Zorko}
\email{andrej.zorko@ijs.si}
\affiliation{Jo\v{z}ef Stefan Institute, Jamova c.~39, SI-1000 Ljubljana, Slovenia}
\author{M.\,Pregelj}
\affiliation{Jo\v{z}ef Stefan Institute, Jamova c.~39, SI-1000 Ljubljana, Slovenia}
\author{M.\,Klanj\v{s}ek}
\affiliation{Jo\v{z}ef Stefan Institute, Jamova c.~39, SI-1000 Ljubljana, Slovenia}
\author{M.\,Gomil\v{s}ek}
\affiliation{Jo\v{z}ef Stefan Institute, Jamova c.~39, SI-1000 Ljubljana, Slovenia}
\affiliation{ Centre for Materials Physics, Durham University, South Road, Durham, DH1 3LE, UK}
\author{Z.\,Jagli\v{c}i\'c}
\affiliation{Faculty of Civil and Geodetic Engineering, University of Ljubljana, SI-1000 Ljubljana, Slovenia}
\affiliation{Institute of Mathematics, Physics and Mechanics, SI-1000 Ljubljana, Slovenia}
\author{J.\,S.\,Lord}
\affiliation{ISIS Pulsed Neutron and Muon Source, STFC Rutherford
Appleton Laboratory, Didcot OX11 0QX, UK}
\author{J.\,A.\,T.\,Verezhak}
\affiliation{Laboratory for Muon Spin Spectroscopy, Paul Scherrer Institute, CH-5232 Villigen PSI, Switzerland}
\author{T.\,Shang}
\affiliation{Laboratory for Multiscale Materials Experiments, Paul Scherrer Institute, CH-5232 Villigen PSI, Switzerland}
\author{W.\,Sun}
\affiliation{Fujian Provincial Key Laboratory of Advanced Materials, Department of Materials Science and Engineering, College of Materials, Xiamen University, Xiamen 361005, Fujian Province, People's Republic of China}
\author{J.-X.\,Mi}
\affiliation{Fujian Provincial Key Laboratory of Advanced Materials, Department of Materials Science and Engineering, College of Materials, Xiamen University, Xiamen 361005, Fujian Province, People's Republic of China}

\date{\today}

\begin{abstract}
One of the key questions concerning frustrated lattices that has lately emerged is the role of disorder in inducing spin-liquid-like properties.   
In this context, the quantum kagome antiferromagnets {\YCu}, which has been recently reported as the first geometrically perfect realization of the kagome lattice with negligible magnetic/non-magnetic intersite mixing and a possible quantum-spin-liquid ground state, is of particular interest.
However, contrary to previous conjectures, here we show clear evidence of bulk magnetic ordering in this compound below $T_N=15$\,K by combining bulk magnetization and heat capacity measurements, and local-probe muon spin relaxation measurements. 
The magnetic ordering in this material is rather unconventional in several respects.
Firstly, a crossover regime where the ordered state coexists with the paramagnetic state extends down to $T_N/3$ and, secondly, the fluctuation crossover is shifted far below $T_N$.
{\red Moreover, persistent spin dynamics} that is observed at temperatures as low as $T/T_N=1/300$ could be a sign of emergent excitations of correlated spin-loops or, alternatively, a sign of fragmentation of each magnetic moment into an ordered and a fluctuating part.
\end{abstract}

\maketitle

\section{Introduction}
A two-dimensional quantum kagome antiferromagnet (QKA) with isotropic Heisenberg exchange coupling between nearest-neighboring $S=1/2$ spins has been in the spotlight of condensed matter physics for several years due to its predicted quantum spin-liquid (SL) ground state \cite{norman2016herbertsmithite}.
In this novel state of matter emerging from geometrical frustration, quantum fluctuations suppress traditional long-range magnetic ordering down to zero temperature.
Instead, a disordered and fluctuating, yet strongly quantum-entangled state is established \cite{balents2010spin,imai2016quantum,savary2017quantum,zhou2017quantum}.
Although a consensus about a SL ground state in the QKA was reached a while ago, its exact nature remains controversial to this day \cite{clark2017closing,changlani2018macroscopically}, as both gapped \cite{yan2011spin,depenbrock2012nature,mei2017gapped} and gapless \cite{iqbal2013gapless,he2017signatures,liao2017gapless} SL states appear as  results of different theoretical and numerical approaches.

Ultimately, these theoretical predictions should be confronted by experiments. 
However, all material realizations of the QKA feature some level of perturbation to the idealized Heisenberg case, which may be of fundamental importance  \cite{norman2016herbertsmithite}. 
Structural disorder is a particularly notorious problem, {\red 
as it may result in bond randomness. 
Such randomness is predicted to induce gapless SL-like phases with random arrangements of valence bonds -- random-singlet states -- for different frustrated lattices \cite{watanabe2014quantum, kawamura2014quantum, uematsu2017randomness, kimchi2017valence}.
These states should possess no characteristic energy scale due to the presence of valence bonds beyond nearest neighbors and nearly-free spins.
Bond randomness can affect rearrangements of valence bonds and related propagation of nearly-free spins and is thus crucial for the understanding of low-lying magnetic excitations \cite{li2019rearrangement}.}
All QKA representatives known so far that lack long-range magnetic order suffer from structural disorder. 
This is substantial in the paradigmatic herbertsmithite compound, ZnCu$_3$(OH)$_6$Cl$_2$, where the Cu-Zn intersite mixing amounts to 5-10\% \cite{mendels2010quantum}.
Other QKA spin-liquid candidates, like tondiite MgCu$_3$(OH)$_6$Cl$_2$ \cite{kermarrec2011spin}, Zn-brochantite ZnCu$_3$(OH)$_6$SO$_4$ \cite{li2014gapless}, and Zn-barlowite ZnCu$_3$(OH)$_6$FBr \cite{feng2017gapped} suffer from a very similar amount of magnetic/non-magnetic ion mixing.
Such mixing can in principle be reduced by introducing larger non-magnetic ions, like in the 
cases of CaCu$_3$(OH)$_6$Cl$_2\cdot0.7$H$_2$O \cite{sun2016synthesis} and EuCu$_3$(OH)$_6$Cl$_3$ \cite{puphal2018kagome}, which however both order magnetically.
A similar situation is encountered in Ga$_x$Cu$_{4-x}$(OD)$_6$Cl$_2$, where frozen magnetic moments persist up to the highest substitution level ($x=0.8$) \cite{puphal2018ga}.
The intersite mixing is reduced also in CdCu$_3$(OH)$_6$Cl$_2$, however, there the introduction of large Cd$^{2+}$ ions leads to a distorted kagome lattice \cite{mcqueen2011cdcu3}.
Therefore, new realizations of the QKA model with no disorder, perfect kagome symmetry, and possibly a SL ground state are still eagerly anticipated,
as this would allow a systematic and unambiguous study of the role of other perturbations, e.g., magnetic anisotropy, on an individual basis.

\begin{figure}[t]
\includegraphics[trim = 0mm 0mm 0mm 0mm, clip, width=1\linewidth]{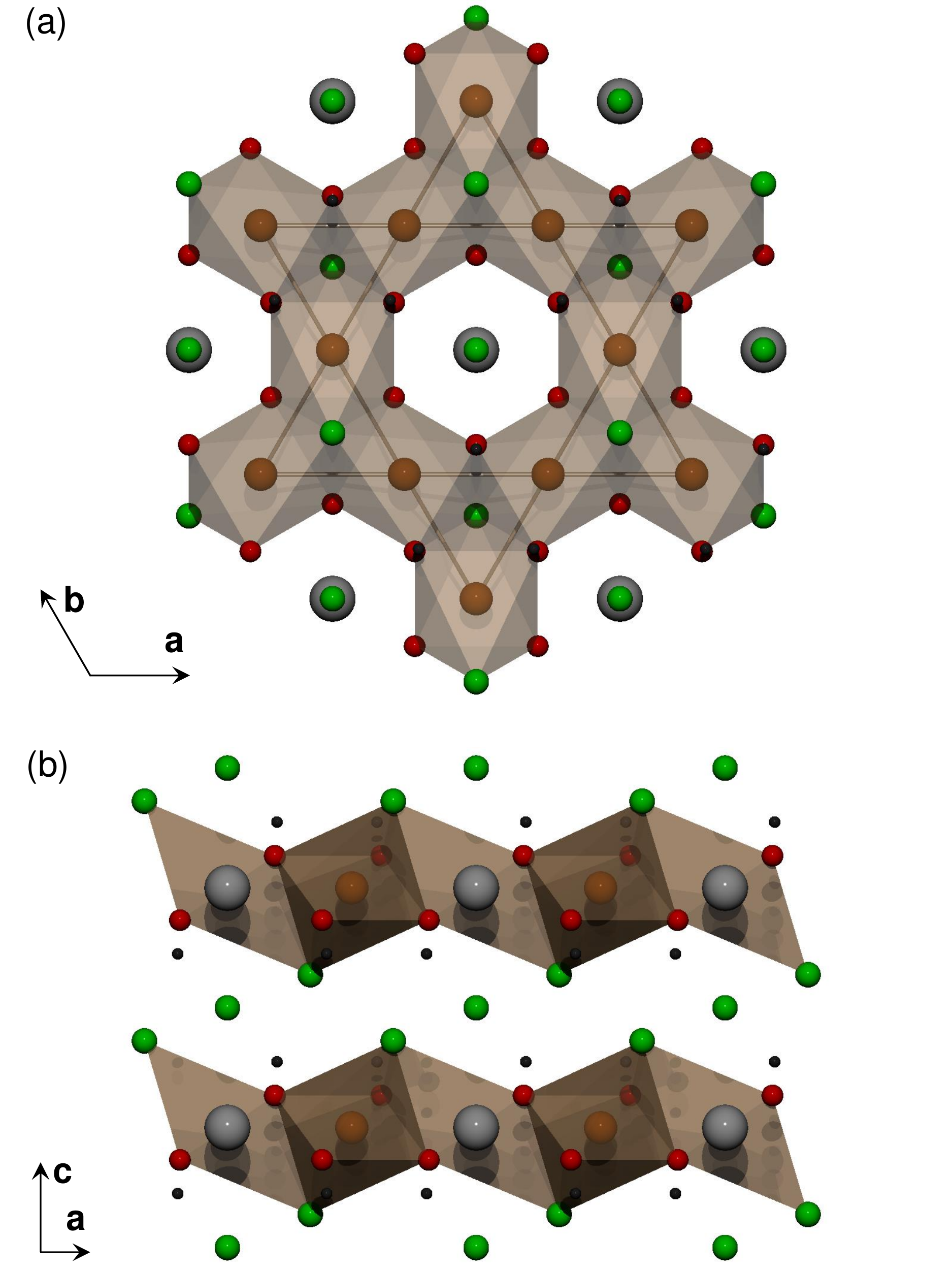}
\caption{A perfect kagome lattice of Cu$^{2+}$ spin-1/2 ions (orange spheres) {\red trapped within O$_4$Cl$_2$ octahedra in \YCu, as view along (a) the $c$ crystallographic axis and (b) the $b$ crystallographic axis. 
The nearest-neighbor exchange interaction (lines) is mediated by OH$^{-}$ groups. 
Y$^{3+}$, O$^{2-}$, H$^{+}$ and Cl$^{-}$ ions are denoted by gray, red, black, and green spheres, respectively.}}
\label{fig1}
\end{figure}

In this regard, we here focus on the recently synthesized yttrium copper chloride hydroxide \cite{sun2016perfect}, {\YCu},
where the bivalent non-magnetic cation Zn$^{2+}$, which is 
positioned between the kagome layers
in herbertsmithite \cite{norman2016herbertsmithite,mendels2010quantum}, is replaced by a much larger trivalent cation Y$^{3+}$,
which resides within the kagome planes.
In \YCu~the arrangement of the magnetic Cu$^{2+}$ ($S=1/2$) ions retains perfect kagome symmetry [Fig.\,\ref{fig1}(a)], as is the case in herbertsmithite, {\red while the kagome planes are well separated by additional chlorine ions that do not belong to the O$_4$Cl$_2$ octahedra around the magentic ions [Fig.\,\ref{fig1}(b)].}
Due to very different ionic sizes of Y$^{3+}$ and Cu$^{2+}$, the intersite
mixing is decreased beyond detectable level \cite{sun2016perfect}. 
Thus by studying the magnetic properties of \YCu~the long-standing issue of the role of defects in QKA could be finally resolved. 
The initial magnetic characterization of \YCu~found sizable antiferromagnetic interactions with the Curie-Weiss temperature of $-99$\,K and proposed that the system should be in a SL state at least down to 2~K, {\red despite a sizable susceptibility increase below 15\,K}  \cite{sun2016perfect}.
Furthermore, no sign of magnetic ordering or freezing was observed in heat-capacity measurements performed between 0.4 and 8~K \cite{puphal2017strong}, making \YCu ~a new promising SL candidate.
However, as bulk measurements can miss more subtle signatures of magnetic instabilities, a local-probe verification of the SL ground state in this compound is required.

In this paper we combine bulk magnetic and heat capacity measurements with local-probe muon spin relaxation ($\mu$SR) measurements.
In contrast to previous claims of a fully fluctuating magnetic ground state \cite{sun2016perfect,puphal2017strong}, our experiments clearly demonstrate the existence of static local magnetic fields below $T_N=15$\,K.
The magnetic ordering is, however, fully established only at temperatures below $T_N/3$.
An additional surprising feature is that the muon spin relaxation rate due to fluctuating internal magnetic fields exhibits a broad maximum well below $T_N$ and remains sizable even in the zero-temperature limit.
This persistent spin dynamics {\red suggest a reduced average static magnetic moments below $T_N$}.

\section{Experimental details}
A high-purity powder sample of \YCu~was prepared according to the procedure published in Ref.\,\onlinecite{sun2016perfect}.
Bulk magnetic characterization was done on a Magnetic Property Measurement System (MPMS) SQUID magnetometer with 261 mg of sample. 
Normalized DC magnetization $M/H$, where $H$ is the applied magnetic field strength, was measured in fields $\mu_0 H=10$\,mT, 100\,mT, and 5\,T between 2 and 300\,K in zero-field-cooled (ZFC) and field-cooled (FC) runs.
AC susceptibility measurements were performed in zero DC field with the driving AC field of 0.6\,mT in ZFC runs over the temperature range 2--30\,K for frequencies between 1\,Hz and 1\,kHz.

Heat capacity was measured with a Physical Property Measurement System (PPMS) instrument in fields of 0 and 9\,T between 2 and 50\,K on a 9.8-mg sample.
The contribution of the addenda was measured separately and subtracted from the data.

$\mu$SR experiments were performed on the MUSR instrument at the ISIS facility, Rutherford Appleton Laboratory, UK, and the GPS instrument at the Paul Scherrer Institute, Switzerland.
Measurements were performed in zero-field (ZF) as well as various longitudinal (LF) and transverse (TF) applied fields with respect to the initial muon polarization.
We used a dilution refrigerator setup on the MUSR instrument to reach temperatures down to 50\,mK and a He-4 cryostat for temperatures up to 21 K.
A few identical temperatures in the range 1.7--4\,K were used on both set-ups to calibrate them.
Approximately 1\,g of powder was fixed on a silver sample holder with a diluted GE varnish to ensure good thermal conductivity.
On the GPS instrument, a standard He-4 setup was used.
The measurements were performed in the veto mode between 1.5 and 20\,K with longitudinal muon polarization.
The same sample from the MUSR experiment mixed with GE varnish was put on a ``fork'' sample holder to minimize the background signal. 
The background signal in the GPS experiment was found to be 2\% of the total signal and was determined from remaining long-time oscillations in TF at 1.5\,K.
The background signal in the MUSR experiment amounted to 10\% of the total signal and was determined from a similar TF run and from comparisons of several ZF and LF runs with runs from GPS measured at identical conditions.
All $\mu$SR measurement in this paper are shown with the background signal subtracted from the original datasets.

\section{Results}
\subsection{Bulk magnetism} 

\begin{figure}[t]
\includegraphics[trim = 0mm 0mm 0mm 0mm, clip, width=1\linewidth]{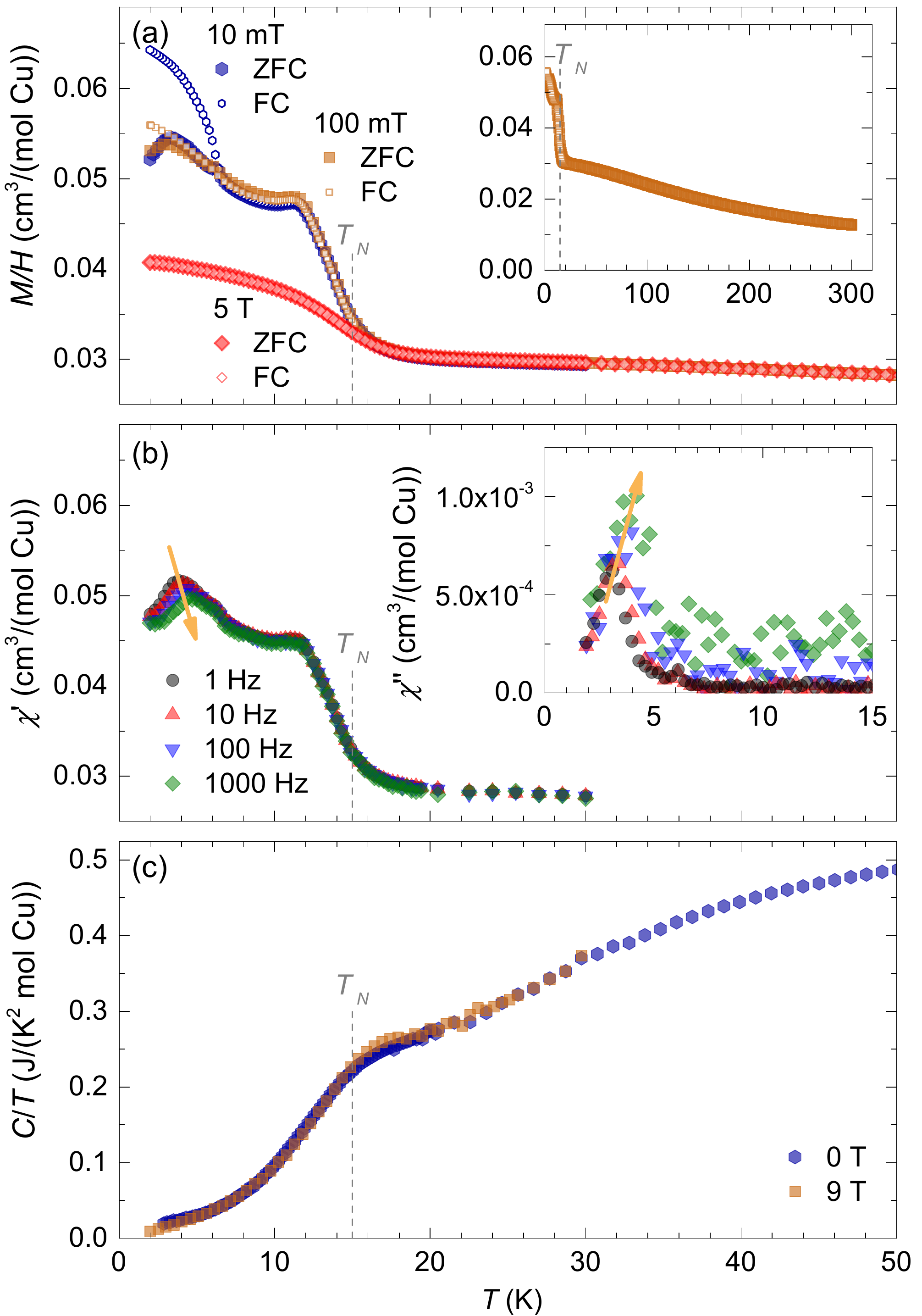}
\caption{(a) DC magnetization of \YCu~normalized by the applied magnetic field.
The inset shows a broader temperature range up to 300\,K and emphasizes the sudden increase of magnetization below $T_N=15$\,K. 
(b) The real part $\chi'$ and the imaginary part $\chi''$ (inset) 
of the AC susceptibility measured in an AC field of 0.6\,mT and zero DC magnetic field. 
The arrows highlight the shift of the $\chi'$ and $\chi''$ maxima with increasing frequency.
(c) {\red The temperature dependence of the heat capacity of \YCu.}
The vertical dashed lines in all panels highlight the ordering temperature $T_N=15$\,K.
}
\label{fig2}
\end{figure}

Bulk magnetic response of our sample in various DC and AC applied magnetic fields [Figs.\,\ref{fig2}(a), (b)] agrees well with published results \cite{sun2016perfect}.
As found previously, the DC magnetization increases steeply below 15\,K in weak magnetic fields, exhibits a hump around 12\,K, and shows a ZFC/FC splitting below 6.5~K [Fig.\,\ref{fig2}(a)].
The latter feature is strongly suppressed when increasing the magnetic field up to 100\,mT, while the other features are unaffected.
Increasing the magnetic field to 5\,T completely suppresses the ZFC/FC splitting and leads to a less intense magnetization increase below 15\,K.
The real part of the AC susceptibility $\chi'$ strongly resembles the DC measurements in low fields [Fig.\,\ref{fig2}(b)].
The susceptibility increase below 15\,K and the hump at 12\,K are frequency independent, while a low-temperature maximum corresponding to the ZFC maximum below the ZFC/FC splitting in the DC measurements, clearly shifts to higher temperatures with increasing frequency;
approximately from $\sim$3\,K to $\sim$4\,K, when increasing the frequency from 1\,Hz to 1\,kHz.
A frequency-dependent maximum at similar temperatures is observed also in the imaginary part of the AC susceptibility [inset in Fig.\,\ref{fig2}(b)].

The ZFC/FC splitting and the frequency dependent maximum in the $\chi'$ are signs of a glassy transition, which hint to a presence of a magnetic impurity.
{\red A tiny fraction 
of clinoatacamite Cu$_4$(OH)$_6$Cl$_2$ that tends to form during the synthesis of {\YCu} \cite{sun2016perfect}, could explain this behavior, as this compound exhibits a spin-glass transition at 6.4\,K \cite{zheng2005unconventional}. 
 In this case, the impurity content can be determined from
the reported saturated FC magnetization of clinoatacamite at 2~K, $M_s\sim0.3$\,Am$^2$/(mol Cu) \cite{zheng2005unconventional}, and the experimental  FC increase of $M\sim0.005$\,cm$^3$/(\,mol\,Cu)$\times H=4\times10^{-4}$\,Am$^2$/(mol Cu) in \YCu~below 6.5\,K in the applied field $\mu_0 H=100$\,mT ($\mu_0$ is the vacuum permeability), where the impurity magnetization is already saturated.}
Correcting the latter by the molar-mass ratio of clinoatacamite and \YCu, $r=0.88$, we estimate a tiny {\red clinoatacamite fraction of only 
 $rM/M_s\sim 0.1\%$.
Alternatively, if the observed ZFC/FC splitting is due to some other spin-1/2 impurity phase, its fraction should be of the same order.}
Except for this low-temperature clinoatacamite contribution, all other features observed in the magnetism of \YCu~are apparently intrinsic.

\subsection{Heat capacity} 

The strong and sudden enhancement of the susceptibility below $T_N$ is, in fact, a signature of a magnetic instability of \YCu, as a clear anomaly is observed also in heat capacity $C$ at the same temperature [Fig.\,\ref{fig2}(c)].
Previous heat capacity measurements by Puphal {\it et al.}~\cite{puphal2017strong} were limited to temperatures below 8\,K and thus missed this feature.
A broad hump that is found around $T_N$ in $C/T$ is field independent
{\red and suggests a broad peak in the magnetic contribution to the heat capacity.}
Both, the magnetization as well as the heat capacity measurements thus strongly suggest a bulk magnetic instability taking place in \YCu~at $T_N=15$\,K that should be attributed to strongly coupled Cu$^{2+}$ spins on the kagome lattice.
We note, however, that these signatures are atypical for an ordinary N\'eel transition to a long-range magnetically ordered state; e.g., a much narrower $\lambda$-type anomaly should occur in heat capacity in this case.  
We also stress that $C/T$ exhibits no obvious feature at 6.5\,K [Fig.\,\ref{fig2}(c)], which
is in line with our suggestion that the ZFC/FC magnetization anomaly and the frequency dependent AC susceptibility peak below this temperature are due to a tiny impurity phase.

\subsection{Muon spin relaxation}

To provide a microscopic insight into the intriguing magnetism of \YCu~we have performed comprehensive $\mu$SR measurements.
Positive muons $\mu^{+}$, being almost 100\% spin polarized when implanted into a sample, are extremely sensitive probes of local magnetism  \cite{yaouanc2011muon}.
The time dependence of their polarization $P(t)$ measured through the spatial asymmetry of emitted positrons at muon decays can be used to determine both, the magnitude and the fluctuation rate of local magnetic fields at the muon stopping site.
Static local fields generally lead to oscillating signals, while monotonic decay of $P(t)$ is found in cases of rapidly fluctuating fields \cite{yaouanc2011muon}. 
A further distinction can be made by applying a longitudinal field (LF), where $P(t)$ gets significantly affected by an applied field of the size of static internal fields, while in the case of fast dynamics of the internal fields $P(t)$ remains essentially unchanged until the applied field by far exceeds the internal fields \cite{yaouanc2011muon}.    

\begin{figure}[t]
\includegraphics[trim = 0mm 2mm 0mm 2mm, clip, width=1\linewidth]{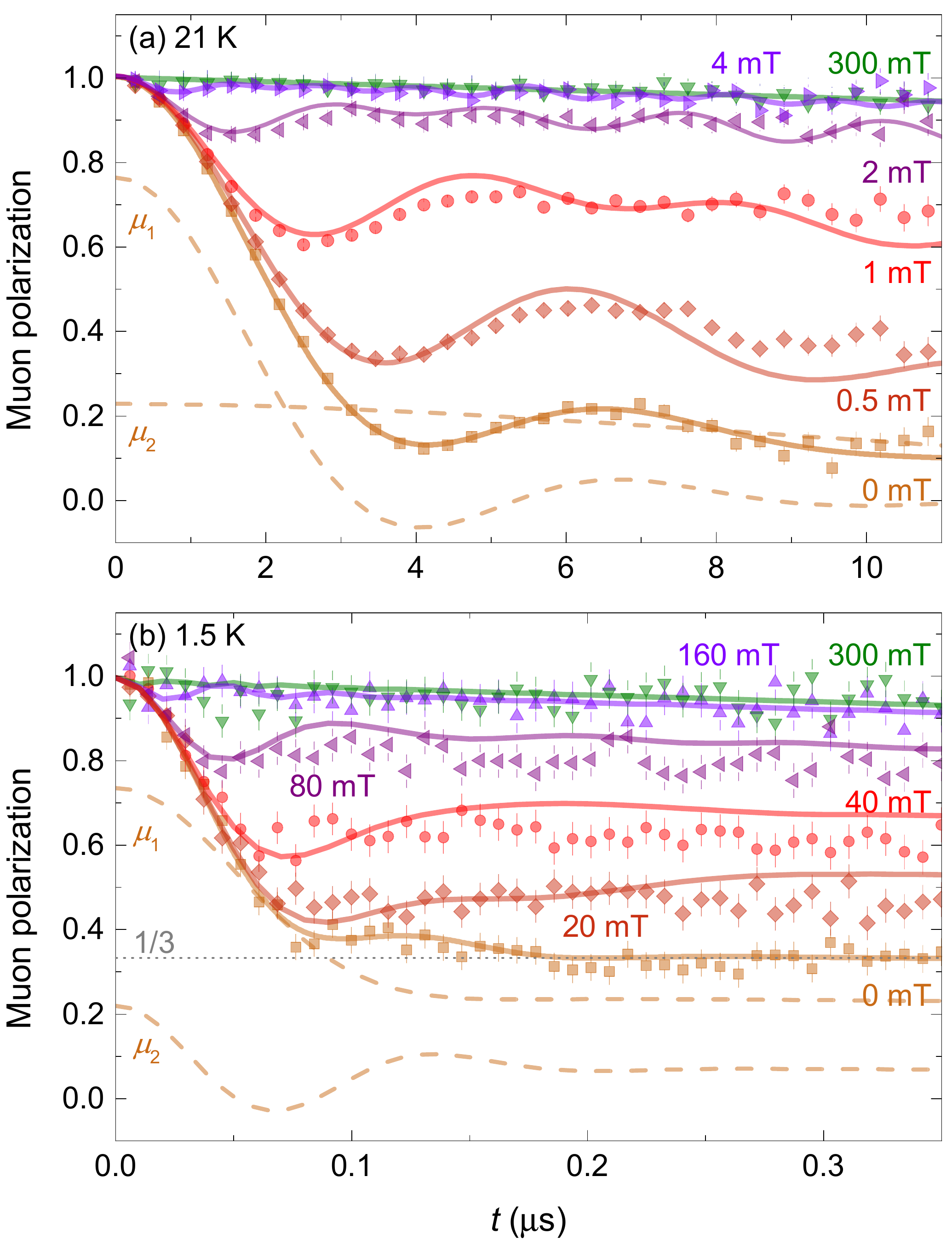}
\caption{The decay of the muon polarization in \YCu~at (a) $T>T_N$  and (b) $T<T_N$ in zero field (ZF) and in various applied longitudinal (LF) magnetic fields (symbols).
The solid lines correspond to two-component fits (shown individually by dashed lines for the ZF datasets).
The majority component $\mu_1$ corresponds to $f=77(3)\%$ of the signal, 
and the minority component $\mu_2$ to $23(3)\%$. 
In (a) the $\mu_1$ component exhibits oscillations due to static nuclear fields in a $\mu$-OH complex, while the decay of the $\mu_2$ component is due to much smaller nuclear fields, yielding the standard Kubo-Toyabe function [the solid lines correspond to the model of Eq.\,(\ref{HT})].
In (b) both components exhibit a damped cosine time dependence [the solid line corresponds to the model of Eq.\,(\ref{LT})].
The dotted line highlights the 1/3 level.
}
\label{fig3}
\end{figure}

At temperatures above $T_N=15$\,K, corresponding to the sudden susceptibility increase and the anomaly in heat capacity, $P(t)$ shows clear oscillations in zero applied field (ZF), as shown in Fig.\,\ref{fig3}(a) for $T=21$\,K.
Applying small longitudinal fields (LF) of the order of 1\,mT already significantly affects the muon polarization, while
$P(t)$ becomes a monotonic, slowly relaxing and field-independent curve for fields above only 4\,mT [Fig.\,\ref{fig3}(a)].
This reveals that the muon polarization is dominantly affected by small static local fields of the order of 1\,mT,
while rapidly fluctuating fields are also present and are responsible for the slow field-independent exponential decay $P(t)={\rm e}^{-\lambda t}$ ($\lambda = 0.015$\,$\mu$s$^{-1}$) that remains present in applied fields $B>4$\,mT.
The small static local fields are due to nuclear magnetism while the dynamical fields originate from electronic magnetism.
The time dependence of the ZF muon polarization at $T>T_N$ is explained by a model containing two nuclear components, 
\begin{align}
P_{\rm HT}^{\rm ZF}(t)=\left[fP_{\rm OH}(t)+(1-f)P_{\rm KT}(t)\right]{\rm e}^{-\lambda t}.
\label{HT}
\end{align}
The first component $P_{\rm OH}$ corresponds to muons that form $\mu$--OH complexes (muon site $\mu_1$), which was previously observed in several kagome compounds containing OH$^-$ groups \cite{mendels2007quantum,kermarrec2011spin,faak2012kapellasite,gomilsek2016instabilities}. The $\mu$--OH complex, in which the muon spin and the hydrogen nuclear spin are strongly entangled, can be modeled in ZF by \cite{gomilsek2016instabilities}
\begin{align}
\nonumber
P_{\rm OH}(t)=\frac{1}{6} &\bigg[ 1+ 2\cos \left(\frac{1}{2}\omega t \right) + \cos \left(\omega t \right) \\
&+ 2 \cos \left(\frac{3}{2}\omega t \right)  \bigg]{\rm e}^{-\sigma t^2},
\label{OH}
\end{align}
where the characteristic frequency $\omega = \frac{\mu_0\hbar}{4\pi}\frac{\gamma_{\rm H}\gamma_\mu}{r^3}$ is a measure of the $\mu$--OH distance $r$. 
Here, $\gamma_{\rm H}$ is the hydrogen gyromagnetic ratio, $\gamma_\mu$ denotes the muon gyromagnetic ratio, and $\hbar$ is the reduced Planck constant.
Additional Gaussian damping ${\rm e}^{-\sigma t^2}$ that is usually observed \cite{mendels2007quantum} accounts for small muon depolarization due to other surrounding nuclei.
The second component $P_{\rm KT}$ in Eq.\,(\ref{HT}) corresponds to muons that sense surrounding static (on the muon time scale) and randomly oriented nuclear fields through a standard Kubo-Toyabe contribution (muon site $\mu_2$) \cite{yaouanc2011muon} 
\begin{align}
P_{\rm KT}(t)=\frac{1}{3}+\frac{2}{3}\left(1-\Delta^2t^2 \right){\rm e}^{-\Delta^2t^2/2},
\label{KT}
\end{align}
where $\Delta/\gamma_\mu$ is the width of the local-field distribution.

After fixing $\lambda$ to the value obtained for high fields, our fit of the 21-K ZF dataset [Fig.\,\ref{fig3}(a)] yields $f=0.77(3)$, $\omega=0.59(2)$\,MHz, $\sigma=0.029(1)$\,$\mu$s$^{-2}$ and $\Delta=0.071(2)$\,$\mu$s$^{-1}$.
We thus find that the majority of muons ($f=77\%$) stop at the expected distance $r=0.159(1)$\,nm from hydrogen of the OH$^{-}$ group \cite{gomilsek2016instabilities}, where the local nuclear magnetic field is of the order of $\omega/\gamma_\mu=0.69$~mT. 
The minority (23\%) muon stopping site $\mu_2$, where the static-nuclear-field-distribution width is $\Delta/\gamma_\mu=0.083$~mT, is most likely in the vicinity of the electrically negative Cl$^-$ ions, as it was also proposed for herbertsmithite \cite{mendels2007quantum} and kapellasite \cite{faak2012kapellasite}, both possessing a very similar chemical formula, ZnCu$_3$(OH)$_6$Cl$_2$, to the investigated compound \YCu.
We note that a single $\lambda$ is used for both muon stopping sites, since it is impossible to determine the two relaxation rates individually at high temperatures due to very slow dynamical muon relaxation.

The field-decoupling experiment, where different longitudinal external fields were applied along the initial muon polarization, was performed to verify that the $\mu$SR polarization at 21\,K was indeed mostly due to small static internal fields.
The muon polarization curves corresponding to $\mu$--OH complexes (site $\mu_1$) in finite LFs were calculated numerically by averaging over many random orientations of a LF with respect to the $\mu$--OH bond, each time diagonalizing the total Hamiltonian including the dipolar coupling of the muon spin with the hydrogen nuclear spin and the Zeeman term.
The Kubo-Toyabe contribution for the $\mu_2$ site in LFs was solved analytically. 
All the parameters were fixed to ZF values, except the Gaussian damping $\sigma$ of the $P_{\rm OH}$ polarization function, since the applied field reduces this damping. 
The calculated curves nicely match the experimental datasets [Fig.\,\ref{fig3}(a)]. 

After establishing full understanding of the $\mu$SR signal at high temperatures, we next inspect if the anomalies observed at $T_N=15$\,K in bulk properties are also seen on a local scale.
Indeed, very different kind of oscillations are found in the ZF $\mu$SR signal at $T<T_N$ as compared to $T>T_N$ [compare Fig.\,\ref{fig3}(a) and (b)].
At 1.5\,K the oscillation frequency is about 40-times faster, meaning that static internal fields at the muon stopping sites are now of the order of several tens of millitesla.
As such, these static local fields can only be of an electronic origin.
The static nature of local fields is again confirmed by measurements in longitudinal fields, as an applied field of a few tens of millitesla significantly affects the $P(t)$ curve and a 160-mT LF completely suppresses the oscillations [Fig.\,\ref{fig3}(b)].
The remaining field-independent monotonic decay of the muon polarization is again due to a dynamical local-field component.
Due to the existence of two muon stopping sites, the ZF low-temperature signal is fitted using a two-components model assuming magnetic order \cite{yaouanc2011muon},
\begin{align}
\nonumber
P_{\rm LT}^{\rm ZF}(t)&= f \left[ \frac{1}{3}+\frac{2}{3}  \cos \left( \gamma_\mu B_1 t \right) {\rm e}^{-(\lambda_{T,1} t)^2} \right]{\rm e}^{-\lambda_{L,1} t} \\
&+(1-f)\left[ \frac{1}{3}+\frac{2}{3} \cos \left( \gamma_\mu B_2 t \right) {\rm e}^{-(\lambda_{T,2} t)^2} \right]  {\rm e}^{-\lambda_{L,2} t},
\label{LT}
\end{align}
where $B_i$ are the two average static internal fields, $\lambda_{T,i}$ the two transverse relaxation rates and $\lambda_{L,i}$ the two longitudinal relaxation rates.
Here, the so-called ``1/3-tail" to which the muon polarization approaches when the oscillations die out [Fig.\,\ref{fig3}(b)], is a characteristic fingerprint of static magnetism in powder samples.
Namely, in a powder sample the projection of initial muon spins on randomly oriented local-field directions yields an average polarization of 1/3 that remains constant \cite{yaouanc2011muon}.
Our fit of the ZF dataset taken at 1.5 K gives the same muon-site occupancy factor $f=0.77(10)$, as determined from the dataset corresponding to 21 K.
We note that a fit with a single component ($f=1$) fails to reproduce the minima in $P(t)$, while adding a third component does not significantly improve the fit.

At 1.5\,K, the two longitudinal relaxation rates are significantly different,
$\lambda_{L,1}=0.05(1)$~$\mu$s$^{-1}$ and $\lambda_{L,2}=0.70(8)$~$\mu$s$^{-1}$, while the transverse relaxation rates are the same within experimental uncertainty, $\lambda_{T,i}=9(1)$~$\mu$s$^{-1}$.
Since $\lambda_{T,i}\gg \lambda_{L,i}$, the former relaxation rates are given by the width of the distribution of static local fields, while the latter are always due to dynamical local fields \cite{yaouanc2011muon}.
The two internal fields differ by a factor of 3.9(5); $B_1 = 13(1)$\,mT and $B_2 = 50(3)$\,mT.
The same parameters also reproduce all LF datasets reasonably well [Fig.\,\ref{fig3}(b)], given the fact that these were modeled with no free parameters by numerically calculated the corresponding static field distributions by averaging over many random orientation of the applied field with respect to the internal field. 

The magnitude of the muon-detected static fields of a few tens of millitesla is typical for spin-1/2 insulators that undergo magnetic ordering.
For example, internal fields of  
60\,mT were found in the ordered state of brochantite, Cu$_4$(OH)$_6$SO$_4$, the parent compound of Zn-brochantite \cite{gomilsek2016instabilities}.
However, in that compound the oscillations are less damped, implying that the relative width of the local-field distribution is much wider in \YCu. 
Here, it is estimated as $\Delta B =\sqrt{2} \lambda_T/\gamma_\mu = 15$\,mT, therefore, it is of the same order as the average magnitude of local fields.
Such damping could either arise in a case of incommensurate magnetic order, or from a broad distribution of muon stopping sites, the latter being a consequence of broad electrostatic-potential minima at the two muon stopping sites.  
Both scenarios allow different relative width of the local-field distributions compared to the average field at the two sites. They can, therefore explain a different shape of these distributions, as revealed by the ratio $\lambda_{T,1}/\lambda_{T,2}$ being about a factor of four larger than the ratio $B_1/B_2$. 
\begin{figure}[t]
\includegraphics[trim = 0mm 25mm 0mm 1mm, clip, width=1\linewidth]{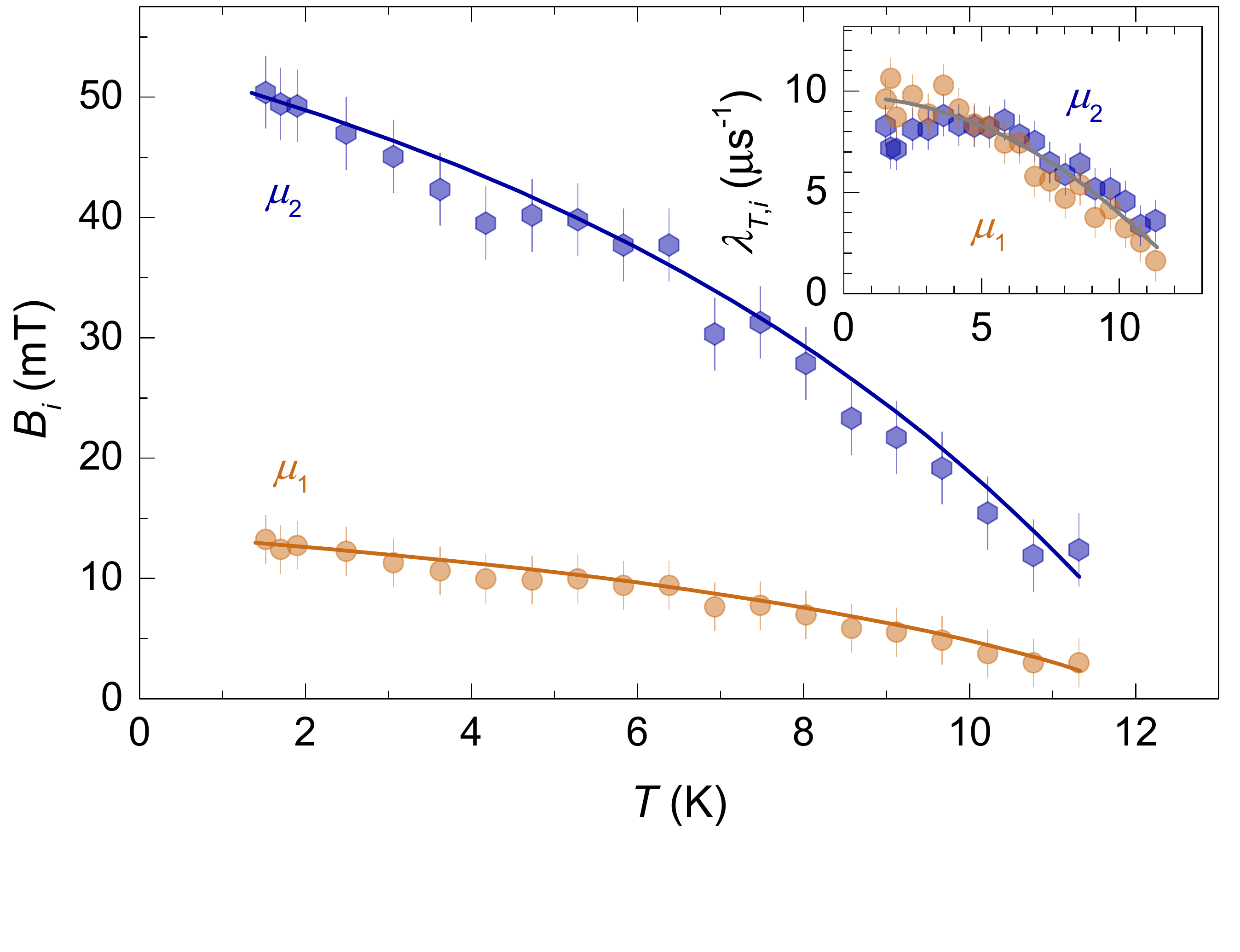}
\caption{
The magnitude of the internal fields and the decay rate of the corresponding oscillations (inset) for the two $\mu$SR components in ZF at $T<T_N$. 
The lines are guides to the eye.
}
\label{fig4}
\end{figure}

The temperature dependence of the internal fields, derived from ZF datasets at different temperatures, reveals the development of a sublattice magnetization, i.e, an order parameter below $T_N$.
The ratio $B_2/B_1 = 3.9(5)$ is temperature independent and can be followed up to about 11\,K  (Fig.\,\ref{fig4}), where the internal fields become too small to be determined reliably.
Both $\mu$SR components thus obviously detect the same magnetic order.
Also the ratio of the transverse relaxation rates $\lambda_{T,2}/\lambda_{T,1} = 1.0(1)$ is temperature independent (inset in Fig.\,\ref{fig4}).

\begin{figure}[t]
\includegraphics[trim = 0mm 11mm 0mm 3mm, clip, width=1\linewidth]{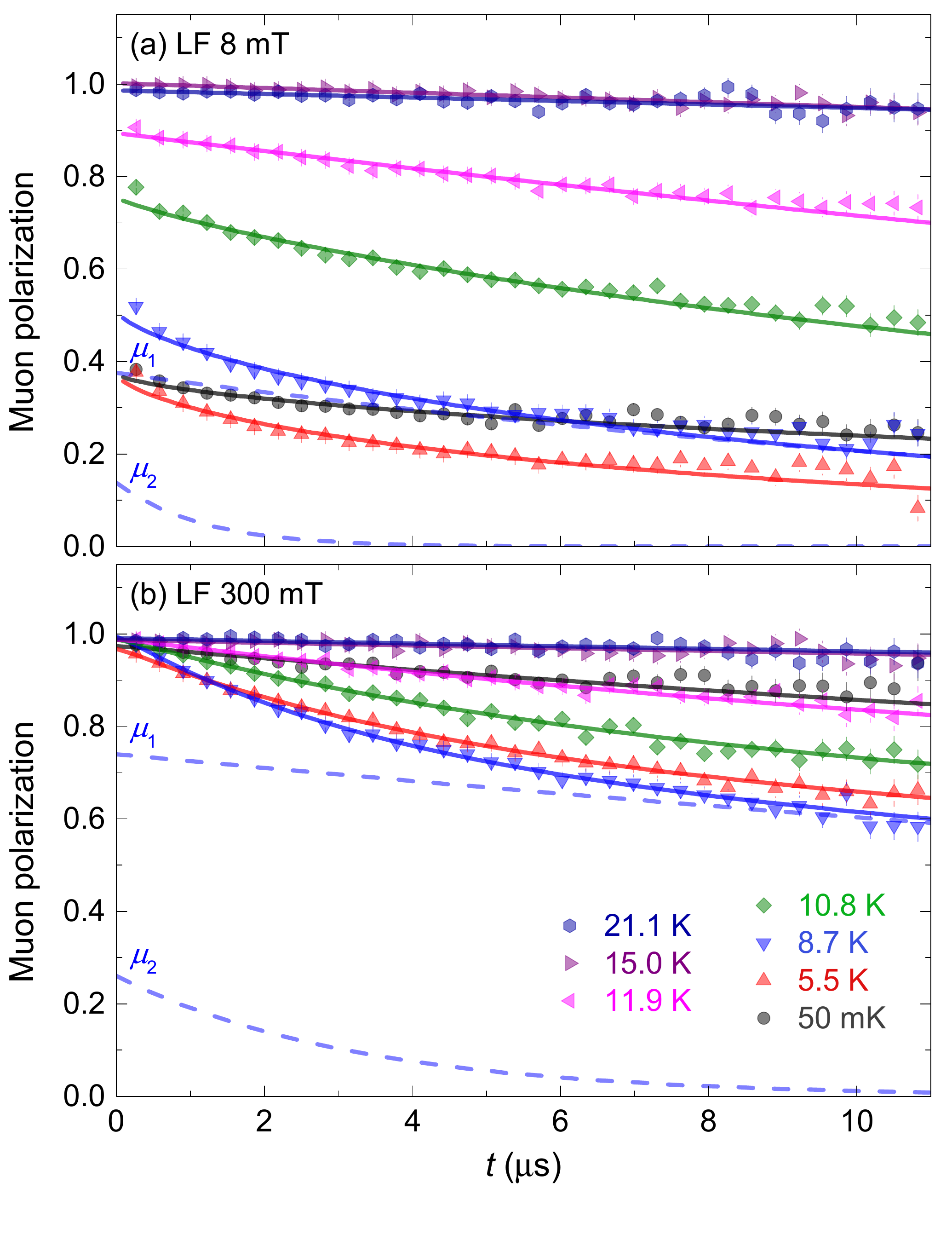}
\caption{The temperature dependence of the muon polarization decay in (a) a weak and (b) a strong longitudinal applied field (symbols) compared to the internal fields. 
The solid lines correspond to a two-component exponential model given by Eq.\,(\ref{LF}).
The individual components are shown for the faster-relaxing datasets (8.7~K) by dashed lines.
}
\label{fig5}
\end{figure}

\begin{figure}[t]
\includegraphics[trim = 0mm 25mm 0mm 1mm, clip, width=1\linewidth]{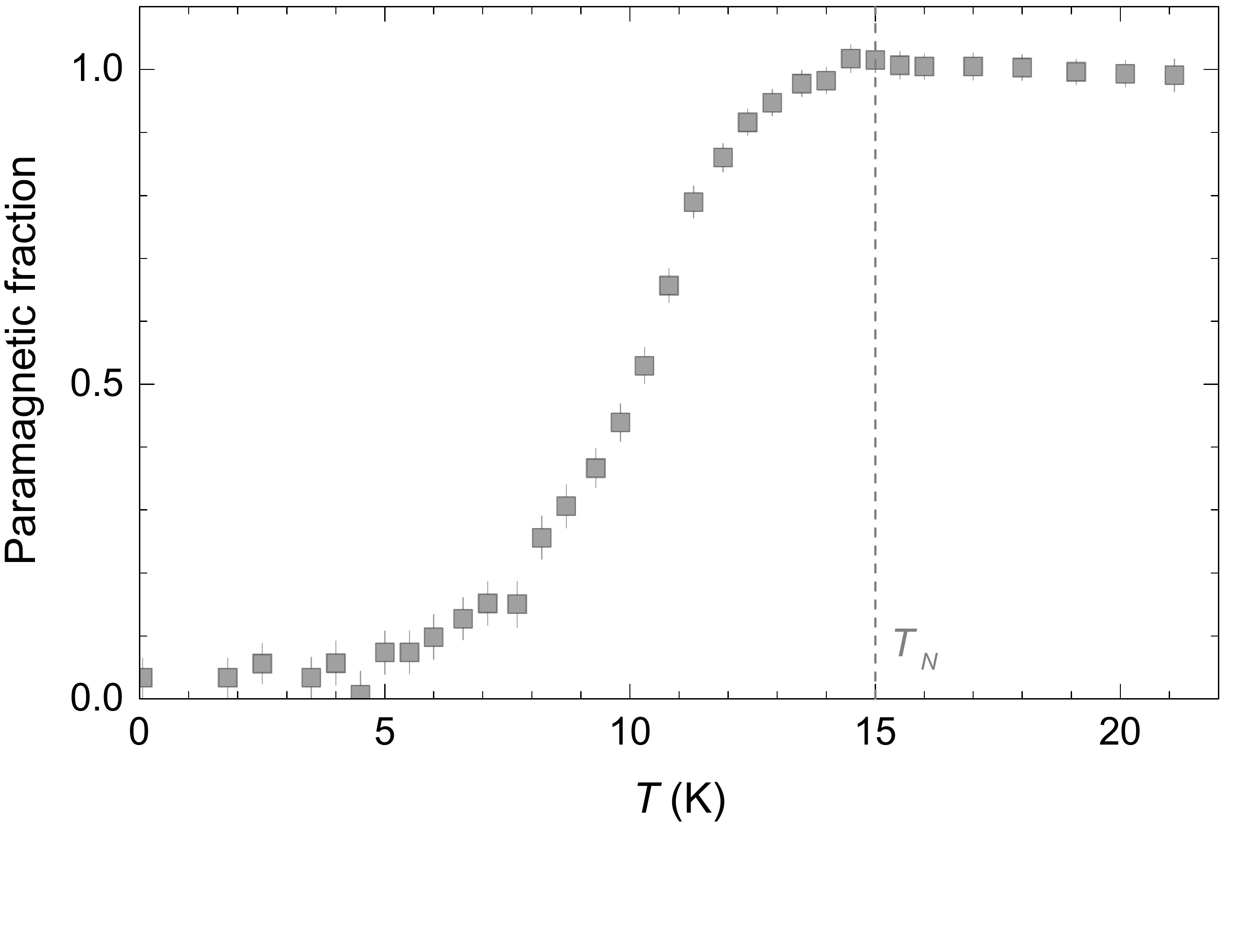}
\caption{
The temperature dependence of the paramagnetic fraction in \YCu, as deduced from the 8-mT datasets (see text for details).
The vertical line denotes the ordering temperature $T_N=15$~K.
}
\label{fig6}
\end{figure}

Contrary to ZF $\mu$SR measurements, applying a small LF of 8\,mT allows us to follow the transition between the paramagnetic and the magnetically ordered state more closely. 
Such a field is large enough to remove muon depolarization due to static nuclear fields, yet small enough not to remove muon depolarization due to the internal fields in the magnetically ordered state.
On an extended time scale compared to the time scale of Fig.\,\ref{fig3}(b) where oscillations due to static internal fields diminish, these oscillations are averaged out [Fig.\,\ref{fig5}(a)].
Therefore, we fitted the dataset taken at 8 mT with the model 
\begin{align}
P^{\rm LF}(t)=P_0(T)\left[ f {\rm e}^{-\lambda_{L,1} t} + (1-f) {\rm e}^{-\lambda_{L,2} t} \right],
\label{LF}
\end{align}
with the fraction $f=0.77$ being fixed from the ZF fits, because it cannot be determined accurately enough from the longitudinal-field datasets alone. 
The two relaxing components at 8.7~K, where the relaxation is the fastest, are individually presented by dashed lines in Fig.\,\ref{fig5}.
The transition into the magnetically ordered state is reflected in a reduced initial muon asymmetry $P_0$.
In the paramagnetic phase $P_0=1$, while in a fully magnetically frozen phase  one should find $P_0=1/3$ in powder samples.
Therefore, the quantity $\eta(T)=(3P_0(T)-1)/2$ serves as a measure of the  paramagnetic fraction at a given temperature.
As expected, $\eta(T)$ starts decreasing below $T_N=15$\,K (Fig.\,\ref{fig6}).
However, surprisingly, the transition into the magnetically ordered state is very gradual, as the paramagnetic part ceases to exist only below $\sim T_N/3 =5$\,K.

Since the oscillations due to static internal fields are averaged out in the 8-mT datasets in Fig.\,\ref{fig5}(a), the relaxation of these datasets is entirely due to dynamics of the local fields.
We find that the relaxation-rates ratio $\lambda_{L,2}/\lambda_{L,1}=15(1)$ is constant for all temperatures below $T_N$ (Fig.\,\ref{fig7}) and exactly matches the ratio of squares of the static internal fields in the ordered state, $(B_2/B_1)^2=15(4)$.
The $\lambda_L\propto B^2$ scaling is characteristic of a fast fluctuation limit, where $B$ denotes the amplitude of the fluctuating field \cite{yaouanc2011muon}.
Therefore, the ratio of the fluctuating-field amplitude and the static-field amplitude is the same for the two muon stopping sites, implying that both dynamical and static fields originate from the same magnetic centers.
The $\mu$SR experiment thus demonstrates the coexistence of rapidly fluctuating and static local fields in \YCu~below $T_N$. 

Similarly as for the paramagnetic fraction $\eta(T)$, the temperature dependence of $\lambda_{L,i}$ in \YCu~is atypical for a regular magnetic transition, where the muon spin relaxation rate by rule diverges at the transition temperature \cite{yaouanc2011muon}.
Instead, we find that both $\lambda_{L,1}$ and $\lambda_{L,2}$ start increasing below $T_N$ and reach a very broad maximum far below $T_N$, i.e., around 5--7\,K (Fig.\,\ref{fig7}).
In the zero-temperature limit (50\,mK) the relaxation rates remain sizable and
are reduced only by a factor of $\sim$2 from the maximal values.
This finding reveals persistent spin fluctuations \cite{pregelj2012persistent} in the magnetically ordered state of \YCu.
In a much larger LF of 300\,mT, which by far exceeds the internal fields and therefore yields temperature-independent initial muon polarization $P_0=1$ [Fig.\,\ref{fig5}(b)], the muon spin relaxation rates follow similar trends (Fig.\,\ref{fig7}).
Above $T_N$, the relaxation rates are field independent, as expected for the paramagnetic regime.
This changes below $T_N$, where the larger applied field substantially decreases the relaxation rates and shifts the maximum in $\lambda_{L,i}$ closer to $T_N$.
Yet, the two relaxation components yield the same temperature independent relaxation-rate ratio $\lambda_{L,2}/\lambda_{L,1}=15(1)$ as found at 8\,mT.

\section{Discussion}
\begin{figure}[t]
\includegraphics[trim = 0mm 25mm 0mm 1mm, clip, width=1\linewidth]{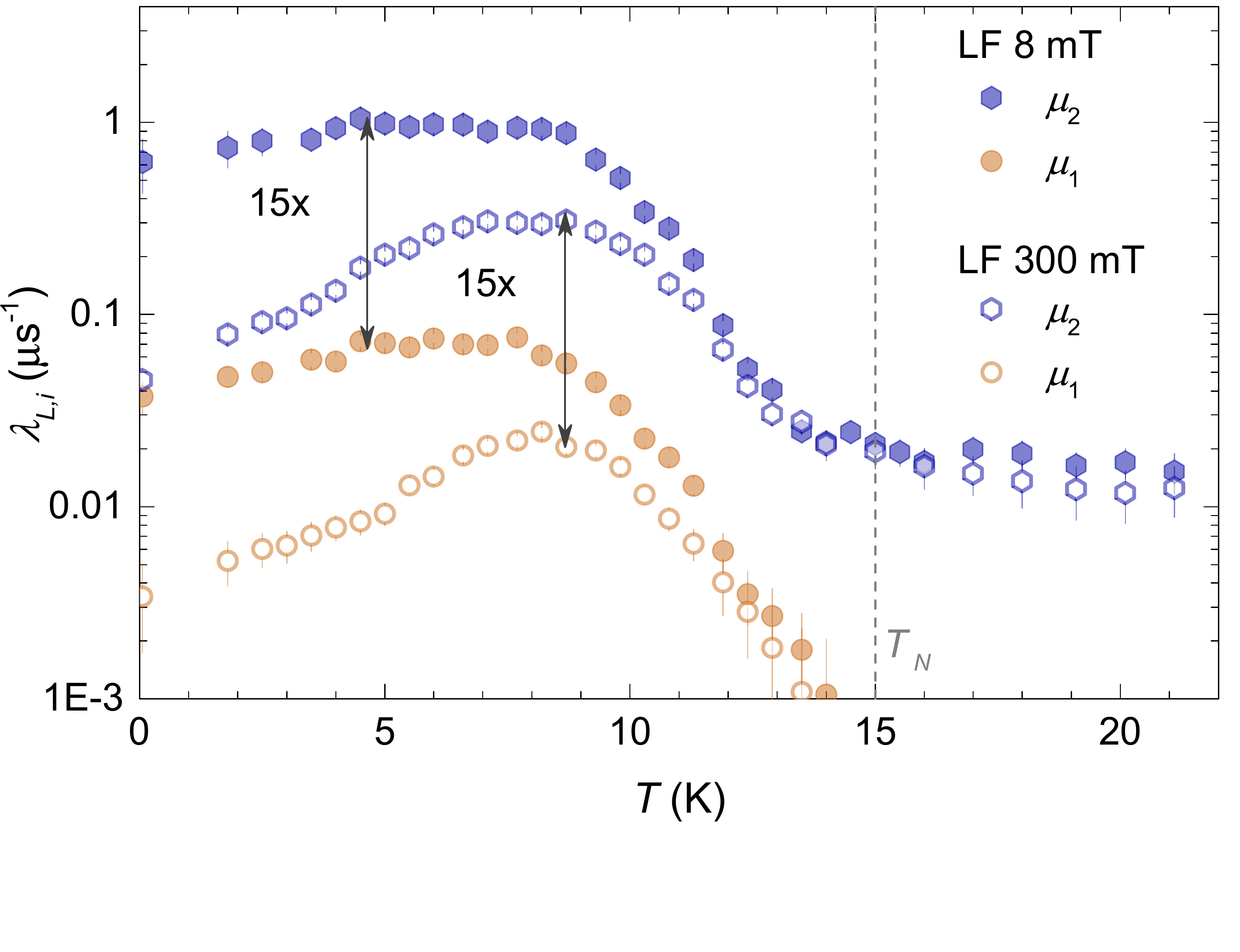}
\caption{
The temperature dependence of the longitudinal muon spin relaxation rate for the two $\mu$SR components $\mu_i$. 
The arrows indicate the ratio $\lambda_{L,2}/\lambda_{L,1}$.
The vertical line denotes the ordering temperature $T_N=15$~K.
}
\label{fig7}
\end{figure}

The combination of bulk magnetization and heat-capacity measurements with local-probe $\mu$SR measurements reported here unambiguously demonstrates that the quantum kagome antiferromagnet \YCu~does not feature a fully-fluctuating spin-liquid ground state, as conjectured in previous studies \cite{sun2016perfect,puphal2017strong}.
Instead, static local magnetic fields of electronic origin are witnessed by $\mu$SR below $T_N=15$\,K [Fig.\,\ref{fig3}(b)].
At this temperature, anomalies are observed also in bulk properties; the magnetic susceptibility shows a sudden increase on decreasing temperature [Fig.\,\ref{fig2}(a)] and the heat capacity exhibits a broad hump [Fig.\,\ref{fig2}(c)].
We note that very similar anomalies in these bulk observables have recently been reported to occur in the structurally equivalent compound EuCu$_3$(OH)$_6$Cl$_3$ at the same temperature \cite{puphal2018kagome}.
 
The magnetic instability of \YCu~is highly unusual in several respects when compared to generic long-range magnetic ordering.
Firstly, the crossover into the magnetically ordered state is very gradual, as the paramagnetic fraction, ceases to exist only at $T_N/3$ (Fig.\,\ref{fig6}).
Such a broad crossover instead of a sharp magnetic transition is consistent with the rather broad maximum observed in the heat capacity [Fig.\,\ref{fig2}(c)].
A similarly broad transition was previously observed in another two-dimensional Cu$^{2+}$-based geometrically frustrated antiferromagnet, CuNCN, where magnetic freezing was found to start below 80~K and to be completed only at 20~K \cite{zorko2011unconventional}.
There, the magnetic state in the crossover regime can be understood as a mixture of paramagnetic and ordered regions on a microscopic scale, which is believed to be a consequence of strong geometrical frustration of the underlying spin lattice \cite{zorko2018magnetic}.
A similar scenario of the coexistence of different phases due to local release of frustration associated with a degenerate ground-state manifold could also apply to \YCu.

Secondly, the longitudinal muon spin relaxation rate does not show critical divergence at $T_N$, nor does it decrease below this temperature, as regularly observed in magnetically ordered phases due to opening of a magnetic-excitation gap.
Instead, both $\lambda_{L,i}$ start increasing below $T_N$ and reach broad maxima at temperatures $T_N/3$--$T_N/2$ depending on the magnetic field (Fig.\,\ref{fig7}). 
A similar fluctuating crossover regime extending far below the transition temperature was observed in the triangular lattice antiferromagnet NaCrO$_2$, where it was attributed to unconventional excitations arising from strong geometrical frustration of the triangular lattice \cite{olariu2006unconventional}.
If a similar scenario of exotic excitations also applies to the investigated kagome antiferromagnet, the corresponding fluctuations of the local fields must be relatively slow, as even a moderate field of 300~mT significantly suppresses the muon spin relaxation at the lowest temperatures.
The fluctuation rate should be of the order of $1/\tau \gtrsim \gamma_\mu B = 250$\,MHz.
We note that the relaxation maximum could also simply be due to a gradual crossover between the paramagnetic and the ordered state when decreasing the temperature below $T_N$.
A maximum in the average relaxation rate below $T_N$ would naturally result from a temperature-independent relaxation of the paramagnetic part and the relaxation of the ordered part due to its collective excitations with a critical divergence at $T_N$.
With decreasing temperature, the former contribution is gradually becoming inferior due to the disappearance of the paramagnetic fraction (Fig.\,\ref{fig6}).

The third intriguing property of the ordered state in \YCu~is that the muon spin relaxation rate remains substantial even at the lowest experimentally accessible temperature, 
i.e., at 50~mK, where $T/T_N=1/300$ (Fig.\,\ref{fig7}).
This is a signature of persistent spin dynamics, a phenomenon often encountered in geometrically frustrated magnets irrespective of the presence \cite{yaouanc2005magnetic, bert2006direct, zheng2006coexisting, pregelj2012persistent,  yaouanc2015evidence, bertin2015nd, xu2016spin} or absence \cite{mendels2007quantum, zorko2008easy, clark2013gapless, gomilsek2016instabilities} of magnetic order.
Two different mechanisms could be responsible for persistent spin dynamics in \YCu.
The dynamics could be a consequence of emergent spin excitations of frustrated magnets related to correlated spin-loop structures \cite{yaouanc2015evidence, bert2006direct}, e.g., spin hexagons in the kagome lattice.
As the dynamics of these spin clusters should be slow due to a large number of spins constituting the cluster \cite{yaouanc2015evidence}, such fluctuations are consistent with the enhanced field dependence of the longitudinal muon spin relaxation rate \cite{bertin2015nd} below $T_N$ (Fig.\,\ref{fig7}). 
On the other hand, the scenario of fragmentation of a single degree of freedom -- the magnetic moment -- into an ordered part and a persistently-fluctuating part is also possible.
This was suggested as
an intrinsic property of amplitude-modulated magnetically ordered states, where magnetic fluctuations were ascribed to the disordered part of the magnetic moment at each magnetic site \cite{pregelj2012persistent}.
Moreover, the concept of spin fragmentation describing the physics of  simultaneously ordered and fluctuating states \cite{brooks2014magnetic} has been recently introduced to spin-ice states on pyrochlore \cite{petit2016observation} and kagome lattices \cite{paddison2016emergent,canals2016fragmentation}.
In such states, strong spin fluctuations have indeed been witnessed through enhanced muon spin relaxation in the ground state \cite{xu2016spin}.
Furthermore, the spin-fragmentation scenario might be a more general property of the kagome lattice with anisotropic exchange interactions \cite{essafi2017generic}.
In \YCu, this scenario of coexistence between partial order and disorder is consistent {\red with the persistent spin dynamics} revealed by muon spin relaxation experiments (Fig.\,\ref{fig7}).

\section{Conclusions}

Despite previous conjectures of a spin-liquid ground state of the novel quantum kagome antiferromagnet {\YCu} with perfect kagome symmetry and no intersite ions disorder, our experiments have clearly disclosed magnetic ordering that arises in this material at $T_N=15$\,K.
This ordering is, however, unusual in several respects; (i) it occurs rather gradually as the paramagnetic fraction ceases to exist only below $T_N/3$, (ii) the muon spin relaxation reaches a broad maximum far below $T_N$ instead of the usual divergence at $T_N$, (iii) the spin dynamics persists to extremely low temperatures, at least of the order of $T/T_N=1/300$.
The persistent spin dynamics {\red implies only partial magnetic order} and could be either due to exotic excitations of correlated spin-loop structures or due to fragmentation of spins into a partially ordered and a disordered part.
The identification of the mechanism leading to magnetic ordering in \YCu~remains an important challenge for future theoretical and experimental studies. 
Importantly, as the effects of intersite disorder and symmetry reduction can be safely dismissed in this compound, the focus can be put on magnetic interactions.
Either magnetic anisotropy\cite{elhajal2002symmetry,zorko2008dzyaloshinsky,cepas2008quantum,zorko2013dzyaloshinsky,
essafi2017generic}, or further-neighbor exchange interactions \cite{messio2011lattice,iqbal2015paramagnetism,gong2015global,buessen2016competing} may play a decisive role in destabilizing the quantum spin-liquid state in \YCu.

{\red
{\it Note added:} After our initial submission, results of another $\mu$SR study of {\YCu} have become available \cite{berthelemy2019local}. The authors reach the same conclusion that long-range magnetic ordering occurs below $T_N$ in the bulk of the sample, although, in their case, a sizable fraction of the sample remains disordered below this temperature.}

\section{Acknowledgements}
This work is partially based on experiments performed at the Swiss Muon Source S$\mu$S, Paul Scherrer Institute, Villigen, Switzerland.
Experiments at the ISIS Neutron and Muon Source were supported by a beamtime allocation RB1720103 from the Science and Technology Facility Council.
These measurements are available at \href{https://doi.org/10.5286/ISIS.E.RB1720103}{https://doi.org/10.5286/ISIS.E.RB1720103}.
The financial support of the Slovenian Research Agency under program No.~P1-0125 is acknowledged.
M.G. is grateful to EPSRC (UK) for financial support (grant No. EP/N024028/1).
The data that support the findings of this study are available via \href{https://doi.org/10.15128/r1n296wz14j}{https://doi.org/10.15128/r1n296wz14j}.
%

%

%
\end{document}